\documentclass[%
 article,
 reprint,
 superscriptaddress,
 showpacs, preprintnumbers,
 amsmath,amssymb,
pre,
 floatfix,
]{revtex4-1}

\usepackage[dvipdfmx]{graphicx}
\usepackage{subfigure}
\usepackage{dcolumn}
\usepackage{bm}
\usepackage{multirow}
\usepackage[mathlines]{lineno}
\usepackage{color}
\usepackage{here}
\usepackage{longtable}

\newcommand{\be}{\begin{equation}}
\newcommand{\ee}{\end{equation}}
\newcommand{\bea}{\begin{eqnarray}}
\newcommand{\eea}{\end{eqnarray}}

\begin{document}

\preprint{ arXiv:1612.07502}

\title{Shannon entropy analysis of the stretched exponential process : application to various shear induced multilamellar vesicles system}

\author{Hirokazu Maruoka}
\email{marcellin1987@gmail.com}
\affiliation{United Graduate School of Agricultural Science, Tokyo University of Agriculture and Technology, 3-5-8 Saiwai-cho, Fuchu-shi, Tokyo 183-8509, Japan}
\author{Akio Nishimura}
\affiliation{United Graduate School of Agricultural Science, Tokyo University of Agriculture and Technology, 3-5-8 Saiwai-cho, Fuchu-shi, Tokyo 183-8509, Japan}
\author{Makoto Yoshida}
\affiliation{United Graduate School of Agricultural Science, Tokyo University of Agriculture and Technology, 3-5-8 Saiwai-cho, Fuchu-shi, Tokyo 183-8509, Japan}
\author{Hideharu Ushiki}
\email{Deceased 7 September 2016}
\affiliation{United Graduate School of Agricultural Science, Tokyo University of Agriculture and Technology, 3-5-8 Saiwai-cho, Fuchu-shi, Tokyo 183-8509, Japan}
\author{Keisuke Hatada}
\email{keisuke.hatada.gm@gmail.com}
\affiliation{Department Chemie, Physikalische Chemie, Ludwig-Maximilians-Universit{\"a}t M{\"u}nchen, Haus E2.033, Butenandtstrasse 11, 81377 M{\"u}nchen, Germany}
\affiliation{Physics Division, School of Science and Technology, Universit\`a di Camerino, via Madonna delle Carceri 9, I-62032 Camerino (MC), Italy}


\date{\today}

\begin{abstract}
The stretched exponential function, $\exp[-(t/\tau_{K})^{\beta}]$, describes various relaxation processes while it has been suggested that the power exponent, $\beta$ is derived from the non-uniformity of the process. In this paper, we attempted to estimate this non-uniformity by introducing Shannon entropy. Shannon entropy evaluates the average information contents of the distribution function, which reflects statistical homogeneity. We investigated the relaxation process of shear induced multilamellar system, which is described with the stretched exponential function. Three types of shear (constant, square, sine) at different frequencies are attempted in order to determine their effects on the relaxation process. We found that the Shannon entropy to which the first moment was introduced is maximized at $\beta$~=~1 : a single exponential. The Shannon entropy of sine shear experiments exhibited the frequency dependence. Thus it is interpreted that the increase of the intensity of shearing and the thermodynamic entropy are reflected on the Shannon entropy. We discussed the meaning of the maximum Shannon entropy in terms of various points of view, it was found that it corresponds to the diffusion process free from the restriction such as geometrical constraints. The constraints and the non-uniformity of process were successfully estimated by Shannon entropy. This study gives a new insight of entropy in general. 
\end{abstract}


\maketitle

\section{Introduction}

The stretched exponential function is defined as follows, 
\begin{equation}
  I\left(t\right)=\exp\left[-\left(\frac{t}{\tau_K}\right)^{\beta}\right]. 
\label{eq:KWW}
\end{equation}
It is widely used in many areas including, glass transition \cite{Majumber}, relaxation process \cite{Yatabe1,Yatabe2}, deformation of real materials \cite{Slonimsky, Oka}, nucleation process \cite{Avrami} and so on. It was discovered by Rudolf Kohlrausch in 1854 \cite{Kohlrausch}, to describe the relaxation process. He showed that the decay of the residual charge of a glass Leyden jar was better described by the stretched exponential function though it was a phenomenological application and he did not provide the microscopic model. Afterwards, his discovery had been almost forgotten until 1970, in which William and Watt reported that the stretched exponential function described the relaxation process \cite{William}. Since its rediscovery, the stretched exponential function has been widely utilized for describing the relaxation phenomena, though its microscopic model has been under debate and is even sometimes considered as a phenomenological tool without any fundamental significance \cite{Phillips}.

\begin{figure*}[t]
\begin{center}
  \includegraphics[width=17.8 cm]{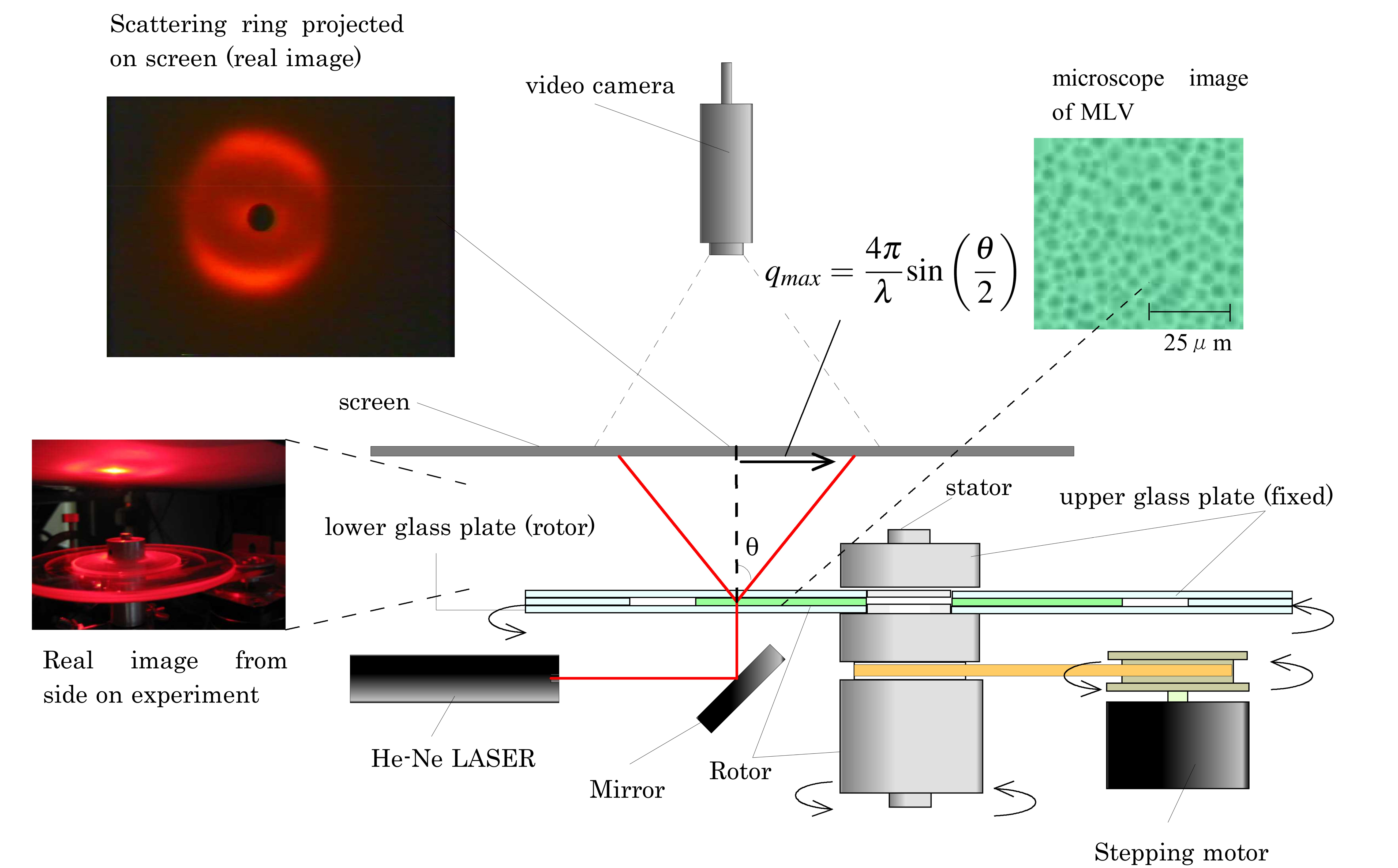} 
\caption{(Color online) The plate-plate type of glass cell was prepared and the sample (Sodium Dodecyl Sulfate/1-Octanol/Brine(20 g/L)) was induced between two glass plates. The upper side of plate was fixed and the lower side of cell was connected with the stepping motor which was controlled by PC, MS-Dos, programmed by Delphi. The experiment was performed at 20 ${}^\circ$C, controlled by air conditioner. Once the program had started, the lower cell was rotated by stepping motor as it was programmed and the sample was sheared. For measurement, the He-Ne LASER (10 mW, $\lambda$=632.8 nm) was radiated to the cell and the scattering light was projected on the screen. The scattering pattern was recorded by camera to VHS recorder, then later the movie is captured and analyzed. By fitting the scattering ring, the radius of scattering ring was measured and we obtain the scattering angle $\theta$. Then we calculate the scattering vector $q_{max}=4\pi/\lambda {\rm sin}\left(\theta/2\right)$.}
 \label{fig:ExpSys}
\end{center}
\end{figure*}

In mathematical form, the stretched exponential function is easily found to be similar to Weibull distribution, which was investigated by Fr\'{e}chet \cite{Frechet}, Fisher and Tippett \cite{Fischer}. Weibull distribution is a third type of the generalized extreme value distribution, and it is also widely used for particle size distribution \cite{Rosin}, extreme value theory \cite{Frechet,Fischer} and failure analysis \cite{Weibull}.

The stretched exponential function is obtained by some microscopic models. Palmer {\it et al.} \cite{Palmer} obtained the stretched exponential function from the model of the hierarchically constrained dynamics in glassy relaxation. Trapping model \cite{Phillips, Shlomo}, the survival time of Brownian motion in disorderly distributed traps, was also one of the major models for the stretched exponential decays. Phillips insisted that molecular relaxation can be understood when one can separate extrinsic and intrinsic effects, and that the intrinsic effects are dominated by two magic numbers, $\beta_{sr}$, short range forces, and $\beta_K$ for long range Coulomb force. These models suggested that the non-uniformity, constraint effects are the essential element to obtain the stretched exponential decays. An interesting result was reported by Bunde {\it et al.} \cite{Bunde}. They associated the power exponent $\beta$ with the scale size effect. They found that the relaxation changes from the stretched exponential to a single exponential decays over a characteristic time which depends on the size of system. This report suggested that the power exponent was a relative, intermediate parameter which depends on the scale.

In this paper, we analyzed the stretched exponential function by introducing Shannon entropy. Several studies reported that the power exponent is highly related with the non-uniformity, constraint effects on the distribution (distribution of spins, sinks etc). Bunde {\it et al.} suggested that the power exponent is not a fixed number, rather it is an intermediate exponent which depends on the size of system. Here we attempts to estimate how the non-uniformity and constraint effect are reflected on the distribution by Shannon entropy. Especially we'd like to focus on $\beta$, as it is the essential parameter to decide the shape of distribution. Shannon entropy is utilized for the probability density function to estimate the average amount of information content of the distribution, which reflects the statistical homogeneity\cite{Shannon}.

The experiment system is the formation process of multilamellar vesicles (MLV) in shear flow. The behavior of the lamellar structures under shear flow has been widely studied \cite{Diat,Herve,Zipfel,Courbin,Leon,Zilman,Yatabe1,Yatabe2}. When the oil/surfactant/water system is sheared, the micro-size, monodisperse vesicles are formed, and the size depends on the shear rate \cite{Diat,Herve}. The vesicles have onion-like structure as it consists of multilamellar. Once it starts to be sheared, the vesicles form and it reaches to the stationary state. The formation mechanism and the behavior of the stationary state are well studied \cite{Diat,Herve,Zipfel,Courbin,Leon,Zilman}. The relaxation process of vesicle size is described with the stretched exponential function \cite{Yatabe1,Yatabe2}. It suggested that the formation process can be described by supposing that the size decreasing is a collective diffusion equation and the initial size distribution is described by the Boltzmann equation \cite{Yatabe1,Yatabe2}. The relaxation process with the shear of the square wave is also described with the stretched exponential function though the power exponent $\beta$ changes depending on the frequency \cite{Yatabe2}. In our work, we tried three types of shears (sine shear, square shear, constant shear) with various frequencies to see how the different way of shearing has impact on the relaxation function and how it is reflected on the Shannon entropy, $H$.

\section{Experiment}
\subsection{Materials}
We prepared a quaternary lyotropic lamellar phase solution by mixing 9 \% Sodium Dodecyl Sulfate (Wako. Co. 97 \% purity), 11 \% 1-Octanol in brine(20 g/L NaCl in distilled water) by weight volume. Solution was mixed to reach the steady state and left for more than two weeks \cite{Yatabe1,Yatabe2}.

\subsection{Measurements}

\begin{figure}[t]
\begin{center}
\subfigure{
 \includegraphics[width=8cm]{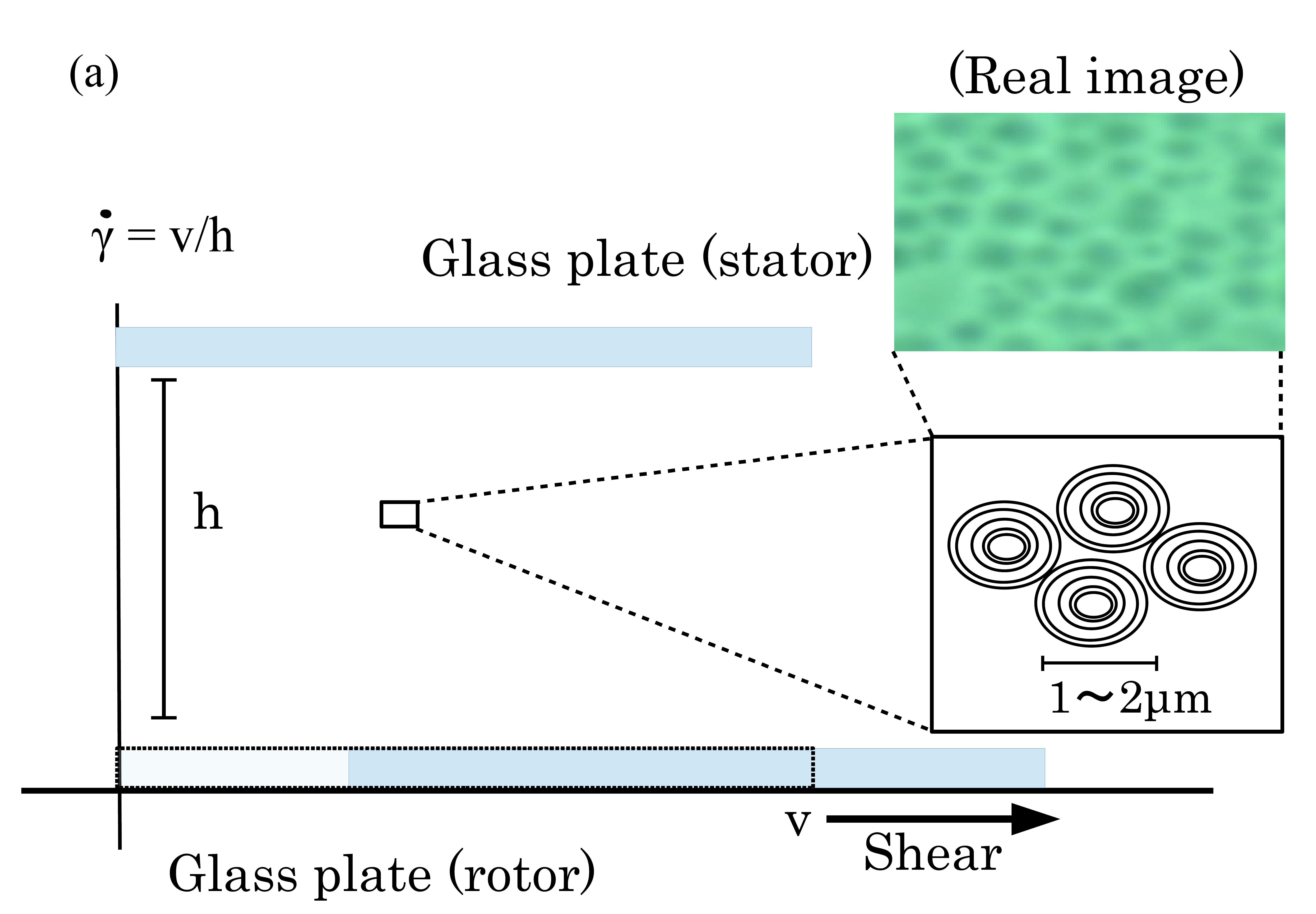}
\label{fig:Fig2B}}
\subfigure{
  \includegraphics[width=8cm]{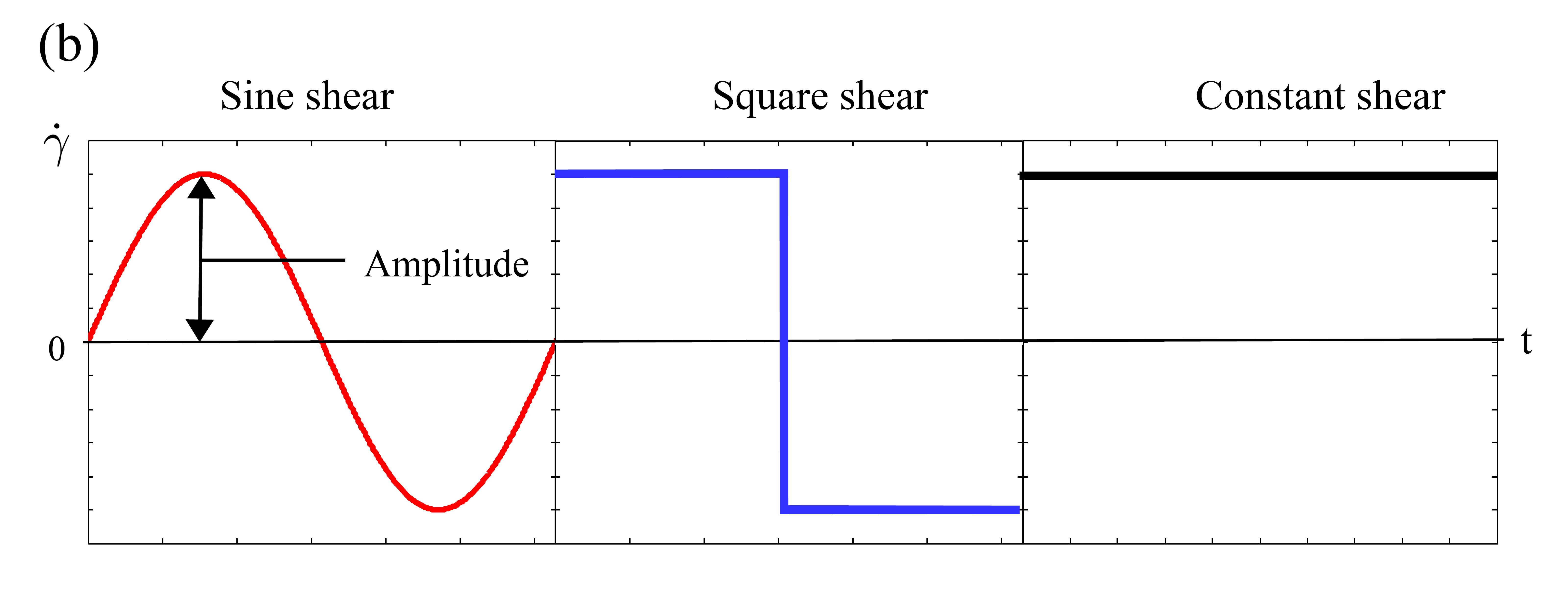}
\label{fig:Fig2}}
\caption{(Color online) (a) The definition of shear rate $\dot\gamma$ : The shear rate, $\dot\gamma$ is defined as the rate at which a progressive shearing deformation is applied.  In the experiment of a plate-plate cell, $\dot\gamma$ is defined as the rotation speed, $v$, divided by the cell gap, $h$ : $\dot\gamma=v/h$. When the mixture of Oil/Surfactant/Water system is sheared, the micro-size, multilamellar vesicles are formed. (b) Variation of shear : various type of shears are prepared, classified with which function shear rate $\dot\gamma$ follows. Sine shear is the shearing of which shear rate varies as the sine function (left). Amplitude of shear is defined as the absolute value of the shear rate in a period, $|\dot\gamma|$. Square shear is the shearing of which shear varies as the square function (middle). In the constant shear, the shear rate and its direction are steady (right). The sine shear and square shear are periodic and the frequency, the number of wave in certain time is defined. We prepared five frequencies :  0 Hz, 6.25$\times$$10^{-2}$ Hz, 3.13$\times$$10^{-2}$ Hz, 1.56$\times$$10^{-2}$ Hz, 7.81$\times$$10^{-3}$ Hz. 0 Hz equals the constant shearing.}
\end{center}
\end{figure}

The experimental system is sketched in Fig.~\ref{fig:ExpSys}. The shear rate is the rate at which a progressive shearing deformation is applied. In the experiment of a plate-plate type cell, the sample is bound between two plates and the shear rate $\dot\gamma$ is defined by the rotation speed of plate $v$ divided by the cell gap $h$ : $\dot\gamma=v/h$ (see Fig. \ref{fig:Fig2B}). We prepared the plate-plate type cell with 1 mm gap between two plates. We inserted the sample between plates. The sample was sheared by rotating lower side of plate. Upper plate was fixed and lower plate was rotated by the stepping motor. Stepping motor was connected with PC (MS-Dos) and we can control the velocity of rotation, the way of shear and frequency. The experiments were performed under 20 ${}^\circ$C with air conditioner. We used the light scattering method for the measurements. He-Ne Laser (10 mW, $ \lambda$=632.8 nm) was directed to the sample and the scattering pattern was projected into the screen. The time evolution of scattering pattern was recorded to VHS recorder with CCD camera. Then we captured the movie to analyze. We measured the scattering angle $\theta$ to calculate the scattering vector $q_{max}=4\pi/\lambda {\rm sin}\left(\theta/2\right)$ by fitting the radius of scattering ring by the software coded by Delphi \cite{Yatabe1, Yatabe2} (see Fig. \ref{fig:ExpSys}). Scattering vector is inversely proportional to the diameter of vesicles, $R \sim 1/q_{max}(t)$. In this experiment, we performed the constant shear,　the square shear and the sine shear which varies the shear rate in accordance with constant velocity, square function and sine function each (see Fig. \ref{fig:Fig2}). We tried different frequencies, 0 Hz, 6.25$\times$$10^{-2}$ Hz, 3.13$\times$$10^{-2}$ Hz, 1.56$\times$$10^{-2}$ Hz, 7.81$\times$$10^{-3}$ Hz. 0 Hz equals the constant shearing. The amplitudes of the shear rates are defined as the absolute values of the shear rate in a period, $|\dot\gamma|$. Strictly speaking, the shear rate is defined as $\dot\gamma=v/h$ when the gradient of the shear is constant such as Couette flow though the exact distribution of the shear is not clear in the sine shear. Therefore $\dot\gamma$ should be considered as the shear rate at wall. In sine shear experiments, the scattering vector fluctuates as sine wave because the shear rate varies as a sine function. Here we collected the plots in the points where shear rate is in the selected amplitude so that we could see the whole time evolution and we could compare with other data.

\section{Result and Discussion}

Before shearing, no scattering vector is observed. The lamellar structures, multil-layered membrane of surfactants, are distributed isotropically. Once the sample was sheared, the scattering ring was observed with small time delay $t_0$, which signified that the multilamellar vesicles had been formed. The diameter of the scattering ring slowly increased and finally reached to the stationary state (see Fig. \ref{fig:Fig3} and \ref{fig:Fig9}). We measured the time evolution of the scattering vector, and it was fitted with the stretched exponential function

\begin{equation}
 q_{max}(t)= q_1+q_2\left[1-\exp\left\{-\left(\frac{t-t_0}{\tau_K}\right)^{\beta}\right\}\right]
\label{eq:FittingEq}
\end{equation} 
by the least-squares method, where $q_1$ indicates the initial scattering vector, $t_0$ is the time delay. The stationary state of scattering vector is $q_1+q_2$. $\tau_K$ is the relaxation time and $\beta$ is the power exponent. All the fitting parameters are shown in Table. \ref{tb:table1} and {\ref{tb:table2}, which are plotted in Fig. \ref{fig:Fig4}.

\begin{figure}[h]
  \includegraphics[width=8.6cm]{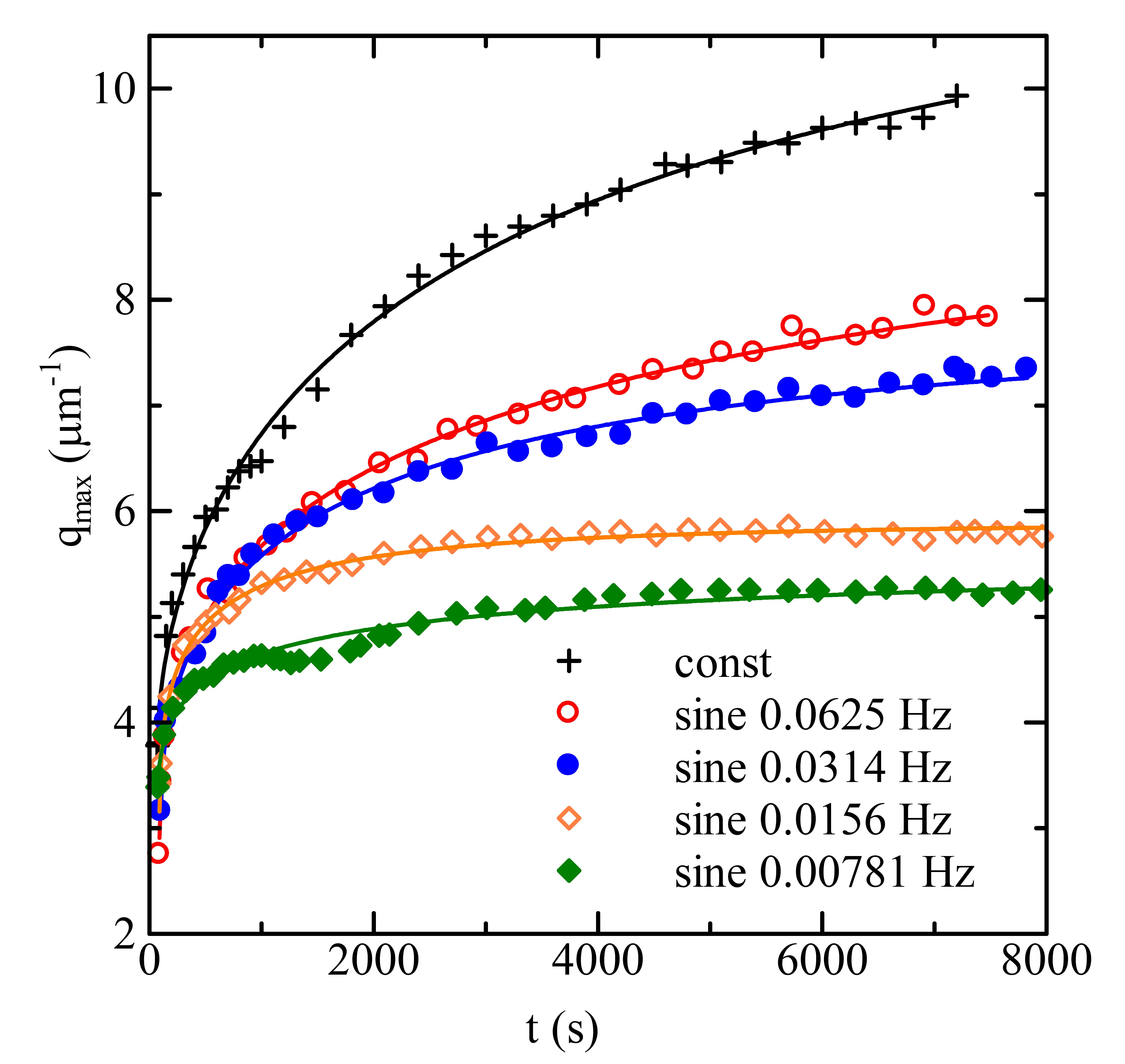}
\caption{(Color online) The relaxation processes of the sine shear at a amplitude of 94 $s^{-1}$ : The shear rates vary following sine function. The experiments were performed under 20 ${}^\circ$C. The amplitudes of sine shear are defined as their absolute values of shear rate, $|\dot\gamma|$. We tried five different frequencies, 0 Hz (+), 6.25$\times$$10^{-2}$ Hz (\textcolor{red}{$\circ$}), 3.13$\times$$10^{-2}$ Hz (\textcolor{blue}{$\bullet$}), 1.56$\times$$10^{-2}$ Hz ({\color{yellow}{$\diamondsuit$}}), 7.81$\times$$10^{-3}$ Hz ({\color{green}{$\blacklozenge$}}). 0 Hz corresponds to the constant shearing. Solid lines are each fitting lines obtained from each plots. }
\label{fig:Fig3}
\end{figure}
\begin{figure}[h]
  \includegraphics[width=8.6cm]{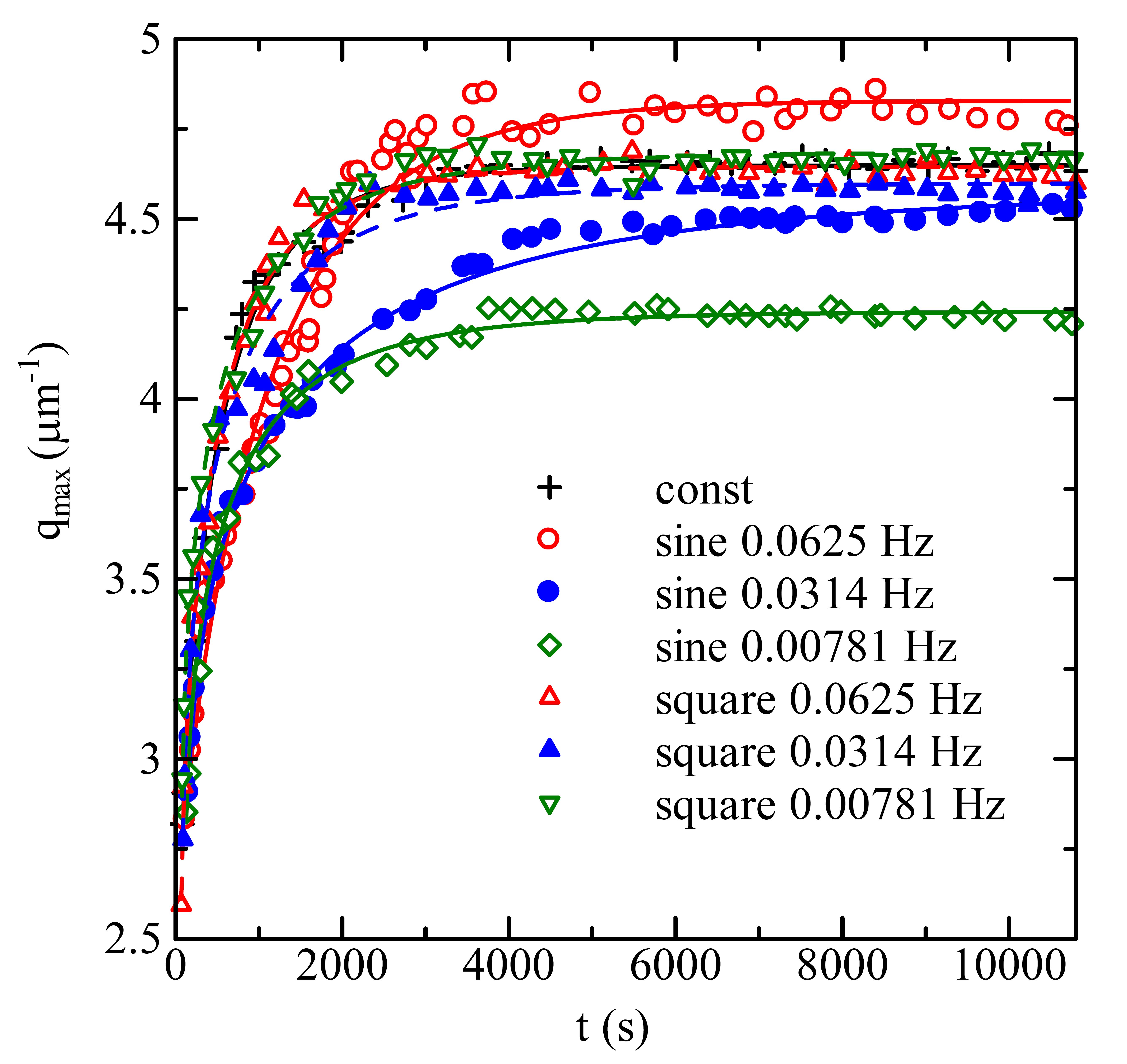}
\caption{(Color online) The relaxation processes of constant, sine and square shears at a amplitude of 70 $s^{-1}$ : The experiments were performed under 20 ${}^\circ$C. constant shear : (0 Hz) (+), sine shear : 6.25$\times$$10^{-2}$ Hz (\textcolor{red}{$\circ$}), 3.13$\times$$10^{-2}$ Hz (\textcolor{blue}{$\bullet$}), 7.81$\times$$10^{-3}$ Hz({\color{green}{$\diamond$}}), square shear : 6.25$\times$$10^{-3}$ Hz (\textcolor{red}{$\vartriangle$}). 3.13$\times$$10^{-2}$ Hz (\textcolor{blue}{$\blacktriangle$}) 7.81$\times$$10^{-3}$ Hz ({\color{green}{$\triangledown$}}). Solid lines are each fitting lines obtained from each plots. }
\label{fig:Fig9}
\end{figure}
\begin{table}[h]
\caption{Fitting parameters, $\beta$ and $\tau_{K}$ with $\chi^{2}$ value obtained from the time evolution of scattering vector $q(t)$ in Fig. \ref{fig:Fig3}. The amplitude is 94 $s^{-1}$. The fitting was performed by least-square method with Eq. (\ref{eq:FittingEq}) \cite{Yatabe1,Yatabe2}. These fitting parameters were used to obtain the distribution in Fig. \ref{fig:Fig5_2} and entropy in Fig. \ref{fig:Fig7}.}
\begin{ruledtabular}
\begin{tabular}{ccccc}
Shear Form & frequency(s) & $\tau_K(s)$ & $\beta$ & $\chi^2$  \\
\hline
constant & 0 & 3469.99 & 0.5673946 & 0.01689165 \\ 
sine &0.00781 & 528.4927 & 0.300365 & 0.007633933 \\
sine & 0.0156 & 363.3605 & 0.4592137 & 0.004646697 \\
sine & 0.0314 & 1584.349 & 0.5039133 & 0.008022212 \\
sine & 0.0625 & 3523.405 & 0.4466833 & 0.008384006 \\ 
\end{tabular}
\end{ruledtabular}
\label{tb:table1}
\end{table}
\begin{table}[h]
\caption{\label{tb:table2}Fitting parameters, $\beta$ and $\tau_{K}$ with $\chi^{2}$ value obtained from the time evolution of scattering vector $q(t)$ in Fig. \ref{fig:Fig9}. The amplitude is 70 $s^{-1}$. The fitting was performed by least-square method with Eq. (\ref{eq:FittingEq}) \cite{Yatabe1,Yatabe2}. These fitting parameters were used to obtain the distribution in Fig. \ref{fig:Fig6_2} and entropy in Fig. \ref{fig:Fig7}.}
\begin{ruledtabular}
\begin{tabular}{ccccc}
Shear Form & frequency(s) & $\tau_K(s)$ & $\beta$ & $\chi^2$  \\
\hline
constant & 0 & 492.8472 & 0.7619915 & 0.001790426 \\ 
sine &0.00781 & 545.3392 & 0.6498283 & 0.00206213 \\
sine & 0.0314 & 992.9321 & 0.5617848 & 0.00131379 \\
sine & 0.0625 & 1087.845 & 0.9137319 & 0.007132354 \\
square & 0.00781 & 458.3162 & 0.6223858 & 0.001871403 \\
square & 0.0314 & 502.9802 & 0.6501096 & 0.003407104 \\
square & 0.0625 & 432.1581 & 0.7149457 & 0.00172249 \\
\end{tabular}
\end{ruledtabular}
\label{tb:table2}
\end{table}

Each plot is obtained from a single experiment. However, we estimated the statistical stability from three sets of experiments in the same condition, and the same set up of the experiment system. At $\dot\gamma$ for $94~s^{-1}$ of constant shear, the results are as follows in the $95\%$ confidence limit : $q_1+q_2$=$10.9 \pm 0.9~\mu m^{-1}$, $\beta$ = $0.46 \pm 0.06$, $\tau_K$=$4015 \pm 938~s$.

In the experiments for 94 $s^{-1}$ of the sine shear (Fig. \ref{fig:Fig3}), we easily find that the relaxation process depends on the frequency. The time evolution of the sine shear approaches to the one of the constant shear with the increase of frequency. The scattering vector of the stationary state $q_1+q_2$ increases with the frequency, Corresponding tendency was seen in the square shear experiment \cite{Yatabe2}. The data for 70 $s^{-1}$ in Fig. \ref{fig:Fig9} does not show as clear a tendency as of 94 $s^{-1}$ in Fig. \ref{fig:Fig3} for $q_1$, $t_0$ and $q_1+q_2$. However in Fig. \ref{fig:Fig4} the relaxation time $\tau_K$ increases with the frequency in the sine shear. In both amplitudes, the fitting parameters of the sine shear show a dependence on frequency.
\begin{figure}[H]
  \includegraphics[width=8cm]{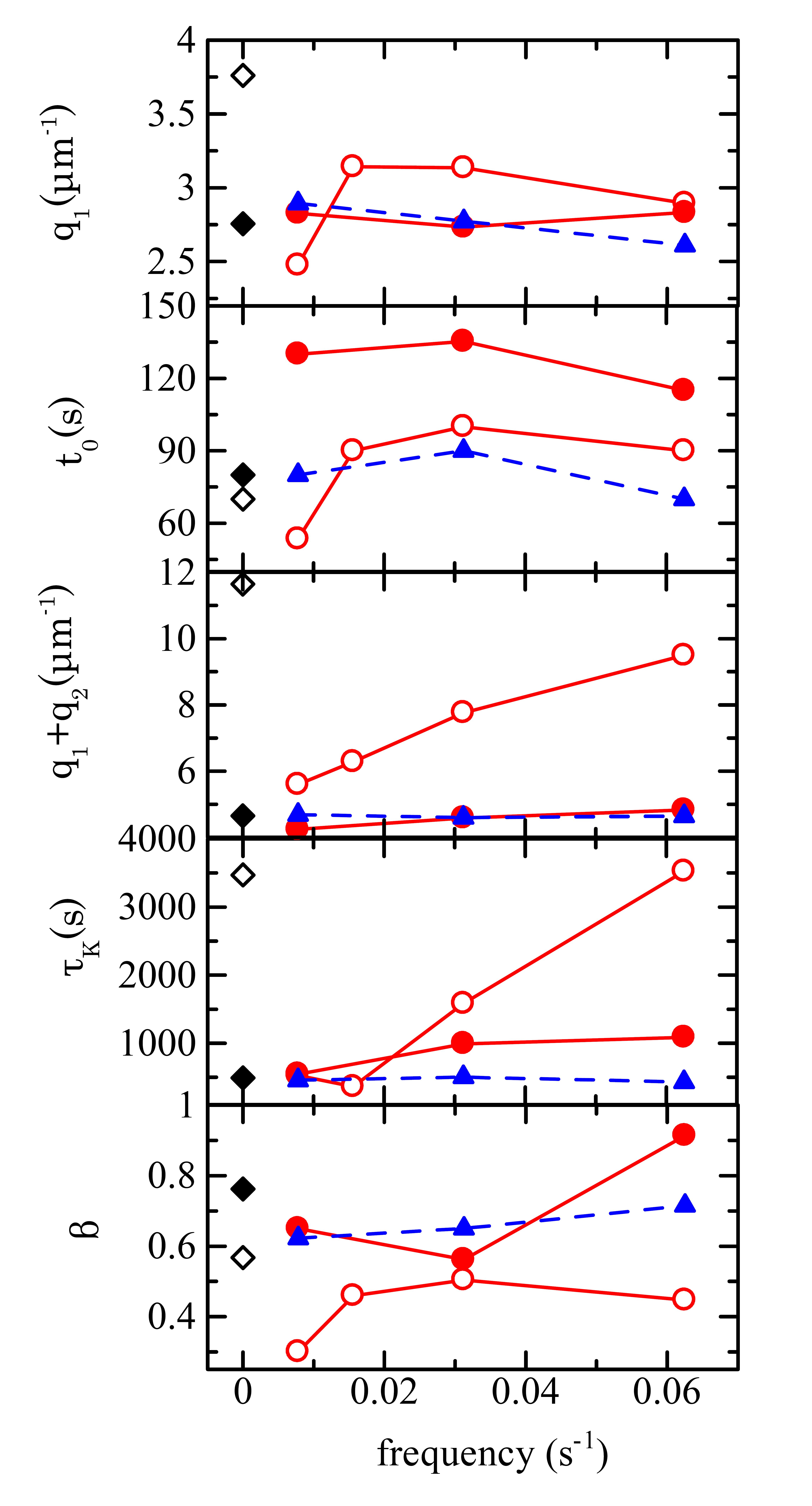}
\caption{(Color online) The plots of fitting parameters of the stretched exponential function Eq. (\ref{eq:FittingEq}) with the frequency of shear. From top to bottom : $q_1$ initial scattering vector, $t_0$ time delay, $q_1+q_2$ scattering vector at stationary state, $\tau_K$ relaxation time, $\beta$ power exponent.  At $|\dot\gamma|$ of 94 $s^{-1}$ : const ($\diamond$), sine ({\color{red}{$\circ$}}). At  $|\dot\gamma|$ of 70 $s^{-1}$ : const ($\blacklozenge$), sine ({\color{red}{$\bullet$}}), square ({\color{blue}{$\blacktriangle$}}).  All parameters were obtained by fitting the plots of time evolution of scattering vector (Fig. \ref{fig:Fig3} and \ref{fig:Fig9}) by the least-squares method. $\beta$, $\tau_K$ are listed on Table \ref{tb:table1} and  \ref{tb:table2} with $\chi^2$ value. In this paper, the data is obtained from a single process. However, we tried to estimate the statistical stability of the experiments in the same condition in the same set up. At $|\dot\gamma|$ of $94~s^{-1}$ at the constant shear, the result were $q_1+q_2$=$10.9 \pm 0.9~\mu m^{-1}$, $\beta$ = $0.46 \pm 0.06$.  $\tau_K$=$4015 \pm 938~s$  in the $95\%$ confidence limit.}
\label{fig:Fig4}
\end{figure}

\subsection{Response functon}

As soon as the shear is switched on, MLV structures form and their vesicle size starts to relax after the time delay. As its relaxation can be considered to be a response from the external force, {\it i.e.}, the dynamical shear, the relaxation function is described with following integral equation,
\begin{equation}
  e^{-\left(\frac{t}{\tau_K}\right)^{\beta}}= \int_0^{\infty} \, D (\tau) \,F (t-\tau)  d \tau
\label{eq:IntegralTrans}
\end{equation}
where $F(t-\tau)$ is the external force and $D(\tau)$ is the response function. Here we consider $F(t-\tau)$ as the step function for simplification. In the stretched exponential relaxation, the following response function is obtained \cite{Slonimsky,Oka, William, Hashimoto},

\begin{equation}
D_{W}(\tau)= \frac{\beta}{\tau_K}\left(\frac{\tau}{\tau_K}\right)^{\beta-1}\exp\left[-\left(\frac{\tau}{\tau_K}\right)^{\beta}\right].
\label{eq:Weibull}
\end{equation} 
Eq.~(\ref{eq:Weibull}) corresponds to Weibull distribution.

Fig. \ref{fig:Fig5_2} and \ref{fig:Fig6_2} illustrate the distribution of the relaxation time, $G(\tau)=\tau D(\tau)$, by using fitting parameters obtained from Table. \ref{tb:table1} and \ref{tb:table2}. While the distribution of lower frequency has broader relaxation time distribution, the higher frequencies have narrower relaxation distribution. This tendency is clearly seen in Fig. \ref{fig:Fig5_2} and it is consistent with the papers \cite{Yatabe2}, which is the case of square shearing fluid experiments. The frequency and the form of shear clearly influence the relaxation process and their distribution.

\begin{figure}[h]
  \includegraphics[width=8.6cm]{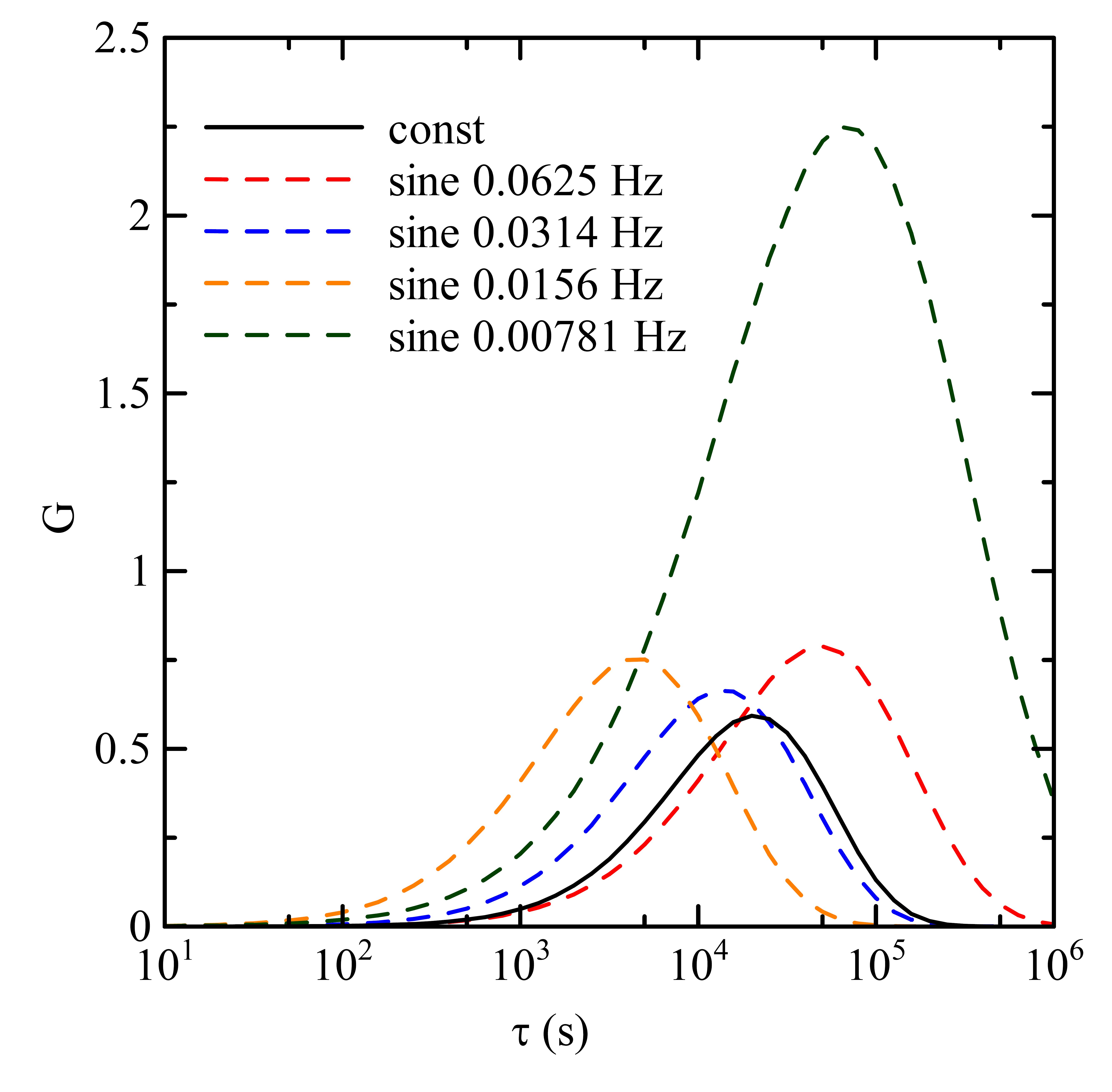}
\caption{(Color online) The Weibull distributions of relaxation times of the relaxation processes in Fig. \ref{fig:Fig3} at a amplitudes of the shear rate of 94 $s^{-1}$. $G(\tau)=\tau D(\tau)$. These were obtained by using the obtained fitting parameters Table \ref{tb:table1}. The frequencies are : 0 Hz (---), 6.25$\times$$10^{-2}$ Hz (\textcolor{red}{- -}), 3.13$\times$$10^{-2}$ Hz (\textcolor{blue}{- -}), 1.56$\times$$10^{-2}$ Hz ({\color{yellow}{- -}}), 7.81$\times$$10^{-3}$ Hz ({\color{green}{- -}}).}
\label{fig:Fig5_2}
\end{figure}
\begin{figure}[h]
  \includegraphics[width=8.6cm]{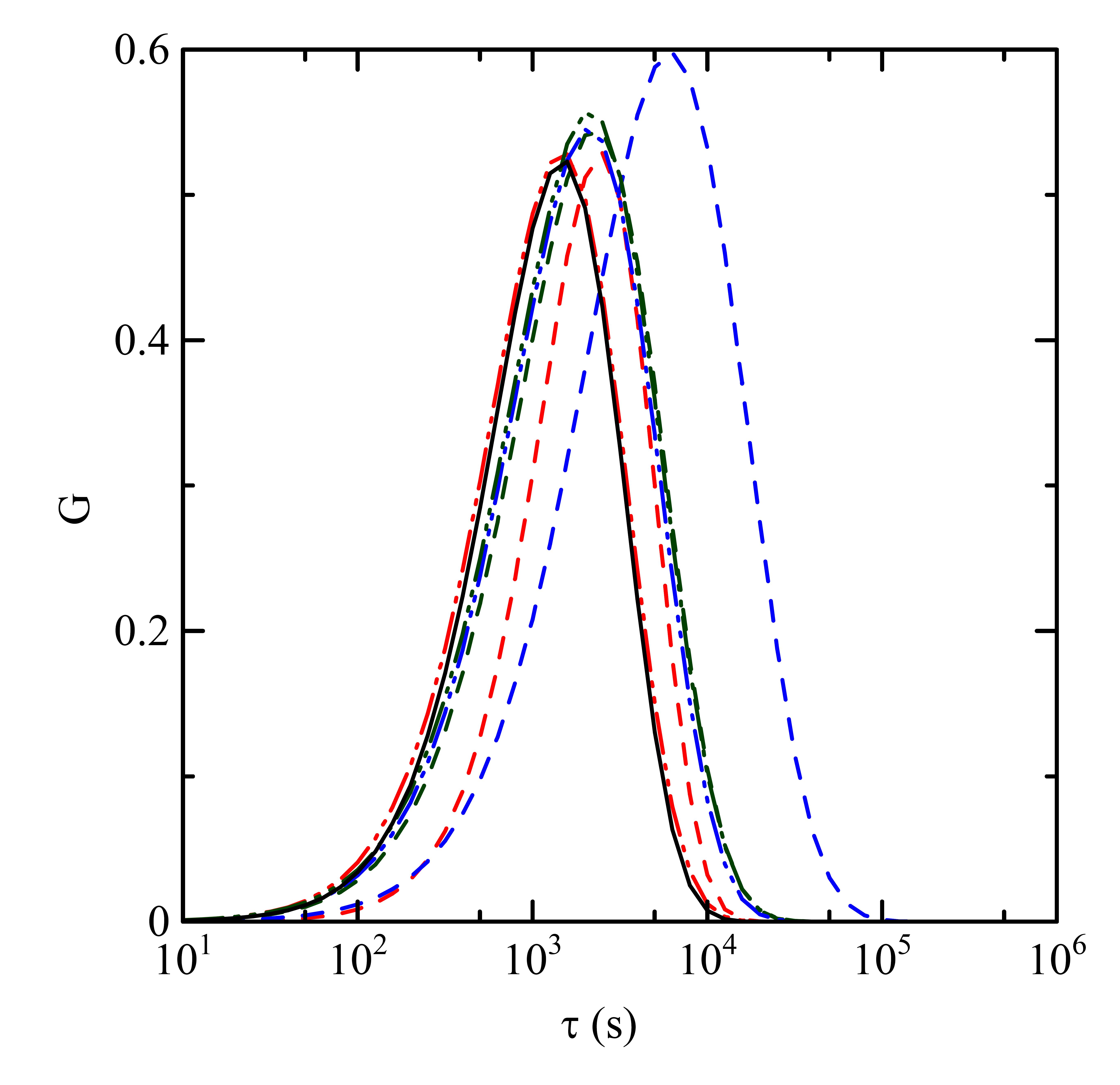}
\caption{(Color online) The Weibull distributions of relaxation times of the relaxation processes in Fig. \ref{fig:Fig9} at a amplitudes of the shear rate of 70 $s^{-1}$. $G(\tau)=\tau D(\tau)$. These were obtained by using the obtained fitting parameters Table  \ref{tb:table2}. The frequencies are : black circle 0 Hz (---), 6.25$\times$$10^{-2}$ Hz (\textcolor{red}{- -}), 3.13$\times$$10^{-2}$ Hz (\textcolor{blue}{- -}), 7.81$\times$$10^{-3}$ Hz ({\color{green}{- -}}). square : 6.25$\times$$10^{-2}$ Hz (\textcolor{red}{$- -\cdot\cdot$}), 3.13$\times$$10^{-2}$ Hz (\textcolor{blue}{$- -\cdot\cdot$}), 7.81$\times$$10^{-3}$ Hz ({\color{green}{$- -\cdot\cdot$}}).}
\label{fig:Fig6_2}
\end{figure}

\subsection{Shannon Entropy}

Now we introduce Shannon entropy, $H$, to evaluate the response function as follows, \cite{Shannon}

\begin{equation}
H=-\int_{0}^{\infty}D(\tau)\ln{\frac{D(\tau)}{C}}d\tau.
\label{eq:Entropy}
\end{equation}
Shannon entropy is the function to evaluate the average amount of information content of the probability density function, which reflects the statistical homogeneity. In this case, we apply the Shannon entropy to $D(\tau)$ of the relaxation process which was obtained from the integral equation in Eq. (\ref{eq:IntegralTrans}) and $H$ evaluates the uniformity of distribution of relaxation time. We can see the condition where the distribution is most homogeneous by obtaining the extreme value of entropy of the response function. $C$ is the dimensional constant so that the function is dimensionless in the Logarithm. The Shannon entropy of Weibull distribution, $D_{W}(\tau)$, is 

\begin{equation}
H_{W}=\gamma\left(1-\frac{1}{\beta}\right)+\ln{\frac{C\tau_K}{\beta}}+1
\label{eq:EntropyWeibull}
\end{equation}
where $\gamma$ is Euler constant. The frequency dependence on Shannon entropy of relaxation processes by using the obtained fitting parameters at Table \ref{tb:table1}, \ref{tb:table2} and the experimental data \cite{Yatabe2} is shown in Fig. \ref{fig:Fig7}. The entropy of sine increases with the frequency though the entropies of square are almost steady as the frequency increase.  

\begin{figure}[h]
  \includegraphics[width=8.6cm]{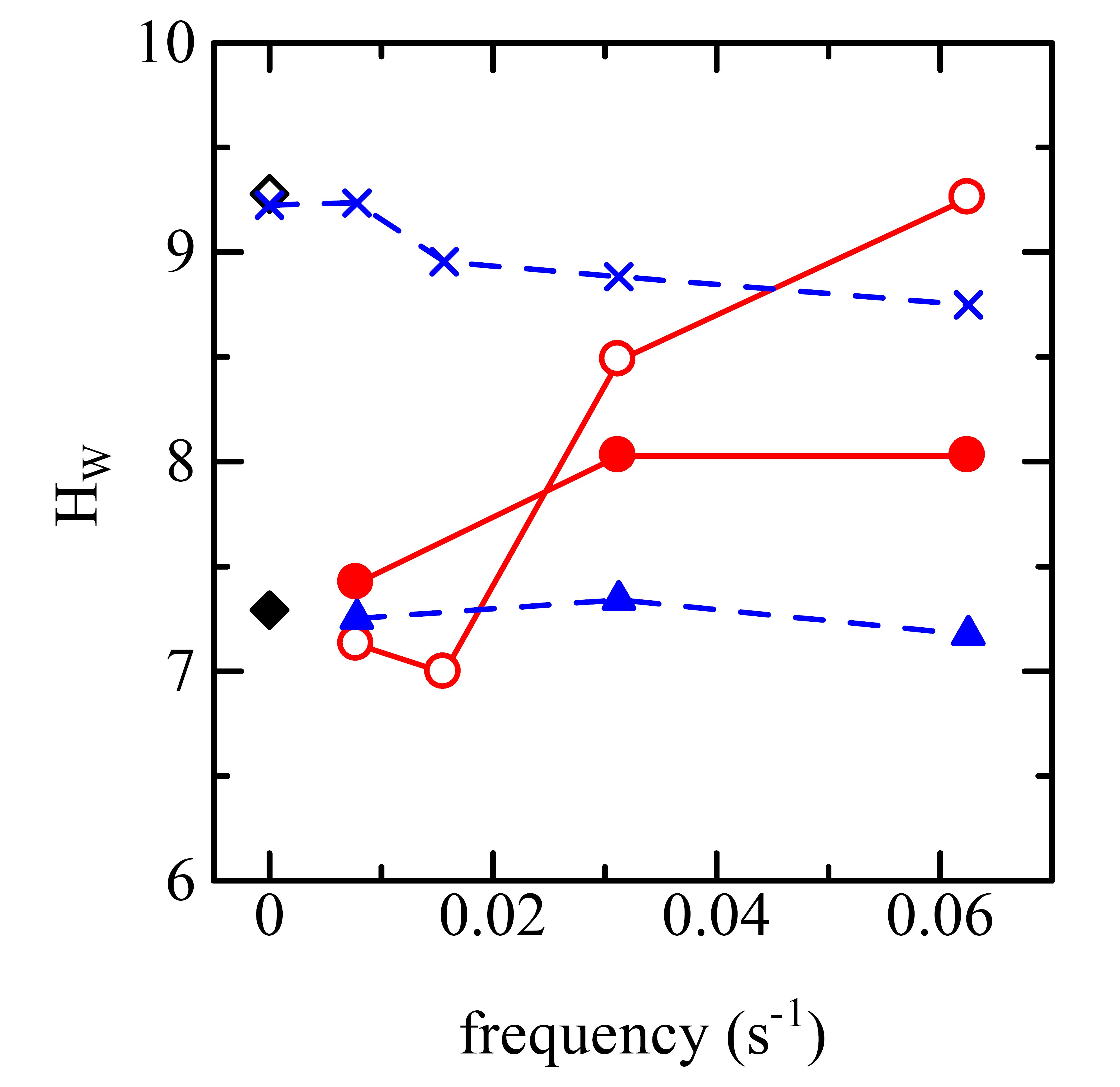}
\caption{(Color online) The frequency dependence of Shannon entropies of Weibull distribution, $D_{W}(\tau)$. The plots were obtained by using the fitting parameter on Table \ref{tb:table1}, \ref{tb:table2} and Table 1 from Ref \cite{Yatabe2}.  At a amplitude of 94 $s^{-1}$ : const ($\diamond$), sine ({\color{red}{$\circ$}}). At a amplitude of 70 $s^{-1}$ : const ($\blacklozenge$), sine ({\color{red}{$\bullet$}}), square ({\color{blue}{$\blacktriangle$}}). At a amplitude of $47~s^{-1}$ : square ({\color{blue}{$\times$}}), which was calculated from the data of Ref \cite{Yatabe2}. }
\label{fig:Fig7}
\end{figure}

Before discussing the result of entropy, let us see the dependence of $\beta$ power component in Shannon entropy. In Eq. (\ref{eq:EntropyWeibull}), it is generally assumed that $\tau_K$ is constant and independent of $\beta$. In such a case, the extreme value is obtained at $\beta=\gamma$ : it is Euler constant \cite{Horst}. However here we think of the case in which the mean value of the relaxation is preserved. The mean value of relaxation time can be obtained by as follows,
\begin{equation}
\langle\tau\rangle=\int_{0}^{\infty} {\tau}D(\tau) d\tau=\frac{\tau_K}{\beta}\Gamma\left(\frac{1}{\beta}\right) \label{eq:MeanRelax}
\end{equation}
where $\Gamma\left(x\right)$ is gamma function. We now introduce Eq. (\ref{eq:MeanRelax}) to the obtained entropy Eq. (\ref{eq:EntropyWeibull})

\begin{equation}
\overline{H}_{W}=\gamma\left(1-\frac{1}{\beta}\right)-\ln{\Gamma\left(\frac{1}{\beta}\right)}+\ln{C}{\langle\tau\rangle+1}. \label{eq:EntropyTau}
\end{equation}
In Eq. (\ref{eq:EntropyTau}), we can easily find the power component $\beta$ in which the entropy is maximized by differentiating the equation with $\beta$ then we obtain,

\begin{equation}
\frac{d\overline{H}_{W}}{d\beta}=\frac{1}{\beta^2}\left[\gamma+\psi\left(\frac{1}{\beta}\right)\right] \label{eq:DiffEntropy}
\end{equation}
where $\psi\left(x\right)$ is digamma function. We easily find that $\overline{H}_W$ is maximum at $\beta$=1 from Eq. (\ref{eq:DiffEntropy}) (see Fig. \ref{fig:Fig8}), which is a condition of a single exponential. It suggests that the stretched exponential is stable at $\beta$=1, approaching to the stationary point. 

\begin{figure}[h]
  \includegraphics[width=8.6cm]{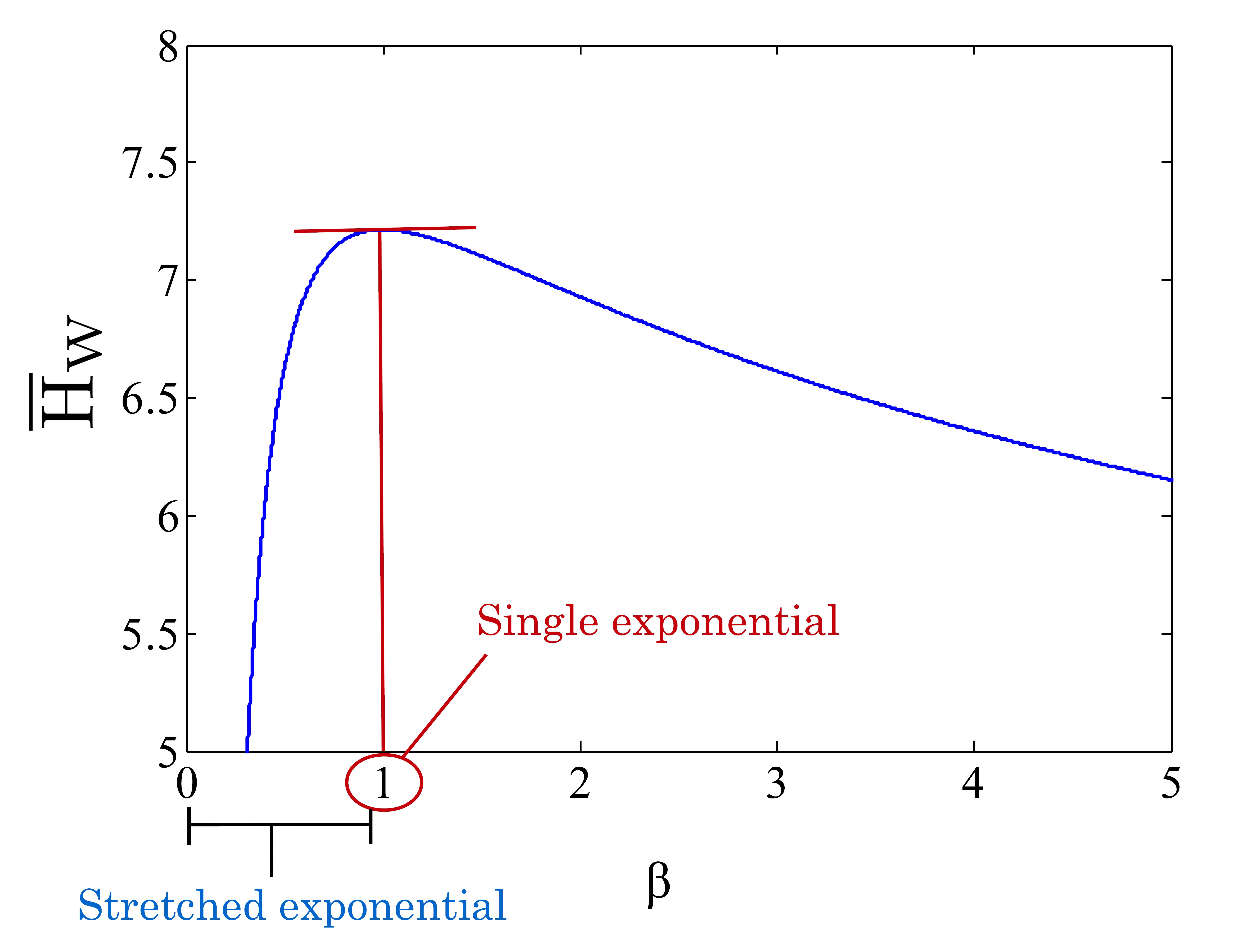}
\caption{(Color online) $\beta$ dependence of the Shannon entropy introduced first moment, $\overline{H}_{W}$, where $C\langle\tau\rangle = 500$. $\overline{\rm{H}}_{\rm{W}}$ is maximized at $\beta=1$ where a single exponential is obtained. }
\label{fig:Fig8}
\end{figure}

We can understand why the entropy of the Weibull distribution is maximized at $\beta=1$, in which the function is a single exponential, when we try to obtain the exponential function with the entropy method \cite{Shannon}. We prepare the functional with the restricting condition of entropy, of normalization and of the preservation of the first moment as follows, 
\begin{equation}
\Phi\left(f\right)=\int_{0}^{\infty}\left[-f\ln{f}+\alpha\left(1-f\right)+\lambda\left(\langle\tau\rangle-\tau f \right)\right]d\tau.
\label{eq:EntExp}
\end{equation}
Obtaining $f$ at a condition of $\delta\Phi(f)=0$ by the variational method, it will be a single exponential function. This is why the entropy which introduced the first moment has the extreme value on $\beta=1$, which is a single exponential, as the preservation of first moment is included in the restricting condition. This result suggests that the Weibull distribution should be considered as a case of $\delta\Phi(f)\neq0$. Luevano reported the same conclusion, the entropy of the stretched exponential probability densities have the maximized condition at $\beta=1$ by introducing the moment though he obtained the stretched exponential probability densities by the entropy method with $\beta$-moment fixed \cite{Luevano}. However we think that the Weibull distribution should be considered as the case when the functional in which first moment fixed is not maximized. 

Now we assume that the single exponential function is the most stable relaxation function in terms of Shannon entropy. When we take this result into account, the relaxation process is more uniform by increasing the frequency in the sine shear experiments. We expect that increasing of frequency corresponds to the increase of the intensity of shearing and it makes the system more disordered. Furthermore as higher frequency has the higher acceleration of shear, the fluid cannot follow the motion of the plate and it elongates the flow, which results in the broader, less laminar velocity distribution of the fluid. It means the increase of the contribution of the viscosity and the flow is more turbulent. Therefore we expect that the energy is dissipated and returned by the heat in the system to increase the thermodynamic entropy in higher frequency. These effects to the thermodynamic entropy is also reflected in the Shannon entropy. The experiments of the square shear and the constant shear are independent of frequency. It is because the acceleration of shear rate is not accompanied in both shear forms and the velocity distribution of fluid is preserved.

\subsection{Interpretation of the condition where the Shannon entropy is maximized.}

Next we interpret the meaning of Eq. (9) by looking other cases where the stretched exponential appears. According to the diffusion controlled process by de Gennes \cite{DeGennes}, the rate constant, $k$, is influenced by its dimensionality. It is time-dependent in the compact diffusion process of kinetics in dense polymer. Such a diffusion controlled process due to the geometrical constraint was pursued by Evesque \cite{Evesque}. He investigated the two molecular reaction of brownian motion in the fractal geometry. If the brownian motion occurs in the fractal geometry, 
the rate equation is described as follows,
\begin{equation}
-\frac{dn_{A}}{dt}=k_{0} t^{\frac{d^{\ast}}{2}-1} n_{A}~n_{B}
\label{eq:Eq_322}
\end{equation}
where $n_A$ and $n_B$ are the concentration of molecular $A$ and $B$, $d^{\ast}$ is a spectral exponent which related with the fractal dimension, $d_f$, of the fractal geometry where the molecules are suspended. Considering a limit $n_B \gg n_A$, the rate equation is described as follows,
\begin{equation}
n_A\left(t\right)\sim\exp\left(-B~t^{\frac{d^{\ast}}{2}}\right)
\label{eq:Eq_323}
\end{equation}
where $B$ = $2k_{0}n_{B}/d^{\ast}$. In this model, the maximized entropy condition corresponds to $d^{\ast}=2$, which equals to noncompact diffusion process or diffusion in two-dimensional Euclidean space ($d^{\ast} = d_f$ = 2). The geometrical constraints are removed in the condition. It is suggestive that the uniformity in the geometry corresponds to the maximum entropy in this case. In the view of diffusion coefficient, as $k \sim D \sim t^{1-\beta}$ \cite{Tsunomori}, the condition of maximum entropy, $\beta=1$, corresponds to the free diffusion. In the trapping model \cite{Phillips,Shlomo}, the stretched exponential is caused by randomly distributed traps. Thus, $\beta=1$ corresponds to the absence of traps. These models suggest that the Shannon entropy reflects the homogeneity of the diffusion process.

In the hierarchical constraint dynamic model \cite{Palmer}, Palmer discussed $\beta$ with the erogodicity. In the strongly interacting systems, these strong interactions are primarily constraints, and it persists over a very wide range of time scales. Therefore, the system is non-ergodic, i.e., equilibrium is partially established when $\tau_0 < t < \tau_{max}$, where $\tau_0$ is a microscopic time and $\tau_{max}$ is ergodic time, which is many orders of magnitude longer. Palmer found the case in which the stretched exponential is obtained in such a non-ergodic condition where a strong constraints persist though a pure exponential is obtained when $t \gg \tau_{max}$. In this case, the maximum entropy condition, $\beta=1$, corresponds to the ergodic relaxation. 

The fact that the entropy which was introduced the first moment, $\overline{H}_{W}$, is maximized at $\beta$=1 where a single exponential function is observed, is quite interesting and consistent with other reports \cite{Bunde,Ogielski,Kakalios}. Bunde {\it et al.} \cite{Bunde} reported that the relaxation function changes from the stretched exponential decays to simple exponential function over the characteristic time. It suggests that the form of relaxation function transforms to a more stable form, from the stretched exponential to the simple exponential. The temperature dependency of the power exponent $\beta$ was reported by Kakalios \cite{Kakalios} through experimental results and was substituted by Ogielski \cite{Ogielski} through simulation results. These results showed that $\beta$ approached to 1 as the temperature increased. These suggest that the thermodynamic entropy is related with the Shannon entropy of the relaxation process as the condition of $\beta=1$ corresponds to the maximized Shannon entropy. In the discussion of the frequency-dependence of entropy in the sine shear experiments, these two entropies are expected to be related. 

In the end, we concluded that the stretched exponential process is an unstable, intermediate process in terms of the entropy. Summarizing all discussion and results including other reports, this instability should be considered as the non-uniformity of the process. Furthermore the stretched exponential is thought to be an intermediate function depending on the scale as Bunde associated it with the scale size, and as Palmer showed that the stretched exponential is obtained in the non-ergodic process. The increase of scale is expected to decrease these non-uniformity relatively. This may be the reason why the stretched exponential process is mostly found in the mesoscopic scale accompanying the inevitable non-uniformity.

\section{Conclusion}

In this paper, we discussed the relaxation process of the multilamellar vesicles in different form of shear (constant, sine and square shear) with various frequencies in which the relaxation process is described with the stretched exponential function. We found that the relaxation process depends on the form of the shear and the frequency and it was reflected on the fitting function. We introduced Shannon entropy to the response function, $D(\tau)$ which we obtained by considering the relaxation process as the integral equation, Eq. (\ref{eq:IntegralTrans}). The Shannon entropy reveals the frequency-dependence of the sine shear process. It is expected that the higher sine shearing accompanies the higher acceleration of the shear rates and it elongates the flow to increase the turbulent. It contributes to the increase of the viscosity and the energy dissipation. 

We found that the Shannon entropy of the Weibull distribution to which the first moment is introduced was maximized at $\beta=1$ where a single exponential was obtained. It was suggested that the stretched exponential should be considered as the unstable function. 

Interpreting this fact by other models, the condition in which the entropy is maximized, {\it i.e.}, $\beta=1$ corresponds to the diffusion free from constraints. In the model of two molecular reactions in fractal geometry, the maximized entropy corresponds to the removal of fractality. In the model of the Trapping model, it corresponds to the absence of traps in diffusion process. 

In the hierarchical dynamical model by Palmer {\it et al.}, the maximized entropy condition corresponds to the ergodic process in which constraints is absent.

Ogielski and Kakalios reported the temperature dependence of $\beta$. Increasing a temperature, $\beta$ converged to 1, in which corresponded to the condition of maximized Shannon entropy. The relation between thermodynamics entropy and the Shannon entropy was suggested. Bunde reported that the stretched exponential changed to the single exponential as the scale increased. It was interpreted that the increasing scale relaxed the non-uniformity relatively. 

These reports and the discussion of our paper suggested that the stretched exponential is non-uniform and constrained process and these non-uniformity was successfully estimated by the Shannon entropy. Shannon entropy can be a useful tool to estimate the non-uniformity of the relaxation process in the mesoscopic scale.

\section{Acknowledgements}

H. M. wishes to thank Masahiko Shoji for advice and discussion of experimental, theoritical design for this paper. The authors wish to acknowledge Tokyo University of Agriculture and Technology Center of Design and Manufacturing, engineering of official of Jun Kinoshita, for preparing some parts of experimental system.



\end{document}